# A Power-Trajectory Framework for Data Center Participation in Day-Ahead Markets

Menghan Zhang and Caisheng Wang
*Department of Electrical and Computer Engineering, Wayne State University, Detroit, MI, USA*
{mhzhang, cwang}@wayne.edu

***Abstract*—The rapid expansion of artificial intelligence (AI) is driving substantial growth in data center electricity demand. Interval-based energy blocks in day-ahead markets do not explicitly retain predictable intra-interval demand evolution, potentially creating mismatches with ramp-feasible generation trajectories and affecting settlement. This paper proposes a power-trajectory framework in which data centers declare predictable aggregate demand evolution through continuous demand-side power trajectories before market clearing. The declared trajectories are treated as exogenous inputs for constructing generation trajectories and the corresponding price trajectory. For feasible market-clearing outcomes, physical temporal consistency is established through continuous-time power balance and satisfaction of the modeled generation operating constraints, while economic temporal consistency is established by preserving the dispatch-price correspondence in energy settlement on a common temporal basis. Under the adopted settlement definitions, a trajectory-based settlement analysis separately quantifies the monetary magnitude of representation-induced power imbalances and the supply-side compensation that may remain under temporally consistent clearing and settlement. An illustrative 5-bus case demonstrates the elimination of representation-induced power imbalance and identifies the timing and physical sources of supply-side compensation. A modified IEEE RTS-GMLC case further illustrates these compensation effects across multiple generating units under workload-driven data center demand variations.**

***Index Terms*—Continuous-time dispatch, data centers, day-ahead markets, power trajectories, trajectory-based settlement.**

## I. Introduction

A recent report indicates that global data center electricity consumption reached hundreds of TWh in 2024 and is projected to grow substantially by 2030 [1]. A North American reliability assessment likewise identifies data centers as an important source of demand growth [2]. Individual data centers can have power capacities of 100 MW or more, while some announced projects are at the gigawatt scale [1]. Compute-intensive workloads can cause substantial temporal variations in power demand [3], and workload coordination has been demonstrated to support grid-interactive operation [4]. For day-ahead market participation, this paper focuses on predictable aggregate demand evolution that can be specified before market clearing, motivating demand representations that retain intra-interval evolution and support demand-generation coordination.

A potential temporal representation gap arises when generation trajectories are coordinated with demand represented only by interval-based energy blocks. Existing generation-scheduling formulations can explicitly represent supply-side temporal characteristics, including continuous-time generation trajectories [5] and startup/shutdown trajectories [6]. However, an interval energy quantity does not, by itself, specify how demand evolves within the interval. A power value assigned to a market interval endpoint does not retain demand variations within that interval. When demand varies within a market interval, matching interval energy quantities or endpoint power levels alone therefore does not guarantee continuous-time power balance between represented demand and ramp-feasible generation trajectories.

This representation issue has both operational and economic implications. Operationally, solely matching interval energy quantities does not necessarily ensure that generation can follow the demand evolution while respecting generation operating constraints [7]. Economically, research on stochastic multi-interval dispatch has shown that intertemporal constraints and pricing schemes can affect dispatch-following incentives and uplift payments [8]. Against this background, this paper examines the settlement effects of temporal representation, including those associated with representation-induced power imbalances and changes in the dispatch-price correspondence. The representation issue here is distinct from forecast uncertainty: even when a demand trajectory is known in advance, reducing it to interval-based energy blocks need not preserve its intra-interval evolution.

Demand response encourages electricity consumers to modify, curtail, or shift their consumption through price-based or incentive-based signals [9]. However, changing electricity consumption and representing its temporal evolution are different tasks. When demand or demand-response schedules are represented only by interval energy quantities, intra-interval demand evolution may still be omitted. Explicit demand-side power trajectories can therefore complement demand-response mechanisms by retaining the intra-interval demand evolution of the corresponding demand schedules.

Load following, regulation, and real-time balancing can accommodate temporal variability and deviations [10]. Explicitly representing predictable demand evolution in day-ahead market clearing can complement, rather than replace, these mechanisms. If known intra-interval demand evolution is omitted, the effects of that omission may be difficult to distinguish from those of residual uncertainty in subsequent balancing. The power-trajectory framework incorporates declared demand-side power trajectories into day-ahead market clearing, while leaving deviations between realized and declared demand to subsequent balancing mechanisms.

One approach is to increase the temporal resolution of day-ahead markets. Higher temporal resolution can reduce approximation errors, but energy quantities assigned to shorter market intervals do not, by themselves, specify intra-interval demand evolution. Previous work [11] has developed power-based market formulations with bidding and pricing rules and also discussed their implications for imbalance settlement. Building on this direction, this paper focuses on a common temporal basis for data center demand representation, day-ahead dispatch, pricing, and settlement.

Building on the operational integration concept in [12], this paper shifts the focus to the role of temporal demand representation in day-ahead market clearing. The trajectory-construction methodologies in [13] and [14] are adopted as enabling tools here. By carrying the declared demand-side power trajectory through generation dispatch, marginal pricing, and settlement on a common temporal basis, the proposed framework makes it possible to examine how a demand-representation choice affects continuous-time physical feasibility and propagates into dispatch-price correspondence and settlement outcomes. This perspective further allows representation-induced power imbalances to be distinguished from supply-side compensation that may remain even under temporally consistent clearing and settlement.

The main contributions of this paper are twofold:

1) A declared demand-side power-trajectory representation is integrated with continuous-time dispatch, pricing, and settlement on a common temporal basis. For feasible market-clearing outcomes, continuous-time power balance and satisfaction of the modeled generation operating constraints establish physical temporal consistency, while preserving the dispatch-price correspondence in energy settlement analysis establishes economic temporal consistency.

2) A trajectory-based settlement analysis is developed to quantify the settlement effects of temporal representation. This analysis evaluates in-market energy payments and supply-side compensation, while separately quantifying the representation-induced monetary mismatch. Unit-level analysis identifies the timing and contributions of minimum-output operation and ramp-constrained transitions, explaining why the compensation may remain under temporally consistent clearing and settlement.

The power-trajectory framework is evaluated using an illustrative 5-bus system and a modified IEEE RTS-GMLC system. The 5-bus case compares four dispatch-price pairings based on interval-based and trajectory-based representations to examine the settlement effects. The modified RTS-GMLC case further illustrates the timing and unit-level distribution of supply-side compensation under workload-driven data center demand variations.

## II. Declared Demand-side Power Trajectory

This section considers two types of data center demand: connection/disconnection-related demand transitions and workload-driven demand variations. These examples illustrate how interval-based demand representations can give rise to representation-induced power imbalances when combined with ramp-feasible generation trajectories. This section then introduces the declared demand-side power trajectory used as an exogenous input to the market-clearing model in Section III.

### *A. Connection/Disconnection-Related Demand Transitions*

This subsection considers planned demand transitions associated with data center connection/disconnection, rather than the instantaneous electrical switching actions. Within this scope, the grid-facing demand transition is assumed to be continuous and specified before day-ahead market clearing. When its intra-interval evolution is omitted from an interval-based demand representation, the transition may instead appear as a step change at a market interval endpoint.

Fig. 1 illustrates this issue for a connection-related demand transition. Let $D(t)$ denote the baseline system demand and $L^{\mathrm{blk}}(t)$ denote the piecewise-constant load profile used in the interval-based demand representation. For illustration, $D(t)$ is held constant over the time range in Fig. 1, including $[\tau_{n-2}, \tau_{n-1}]$ and $[\tau_{n-1}, \tau_n]$. The represented data center demand is zero over $[\tau_{n-1}, \tau_n)$ and increases to $L^{\mathrm{blk}}(\tau_n)>0$ at $\tau_n$. The black step load profile in Fig. 1(a) represents the resulting total demand $D(t)+L^{\mathrm{blk}}(t)$. This step change characterizes the market representation, not an instantaneous change in the underlying physical demand. The illustration is based on power levels at the market interval endpoints $\tau_n$, rather than on an energy-equivalent approximation of a continuous demand trajectory.

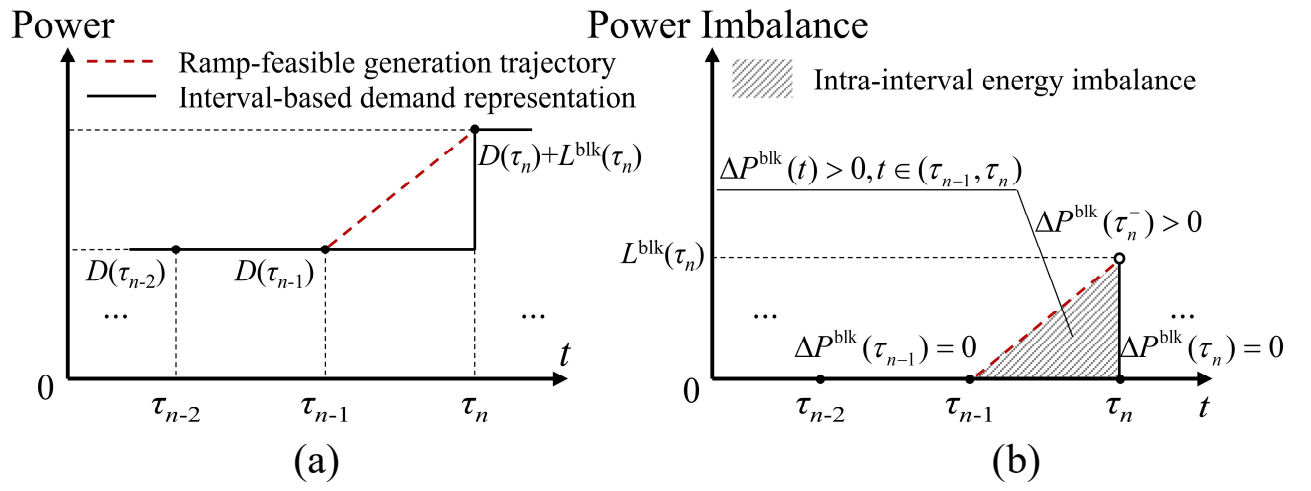

Fig. 1. Representation-induced power imbalance for a planned connection-related demand transition: (a) interval-based demand representation and an illustrative ramp-feasible generation trajectory; (b) the power imbalance $\Delta P^{\mathrm{blk}}(t)$ and the associated intra-interval energy imbalance.

Since the generating units are subject to finite ramping limits, their aggregate output cannot follow this step load profile instantaneously. The red dashed line in Fig. 1(a) illustrates a linear aggregate generation trajectory connecting the total generation power levels at $\tau_{n-1}$ and $\tau_n$. Other connecting aggregate trajectories may be used, provided that the unit-level trajectories satisfy the generation operating constraints. Thus, matching the represented demand at the market interval endpoints does not ensure power balance throughout this interval. The resulting representation-induced power imbalance is defined as follows:

$$\Delta P^{\mathrm{blk}}(t)=\sum_{k=1}^{K} G_k(t)-D(t)-L^{\mathrm{blk}}(t), t\in[\tau_{n-1},\tau_n], \qquad (1)$$

where $G_k(t)$ denotes the generation trajectory of unit $k$, and $K$ is the total number of generating units. Positive and negative values of $\Delta P^{\mathrm{blk}}(t)$ indicate surplus generation and generation shortfall, respectively, relative to the represented total demand.

Equation (1) measures the difference between aggregate generation trajectories and represented demand. The generation trajectory itself remains continuous, while the represented demand undergoes a step change at $\tau_n$. As shown in Fig. 1(b), $\Delta P^{\mathrm{blk}}(t)$ increases during the generation ramp over $[\tau_{n-1}, \tau_n)$ and drops to zero at $\tau_n$ as the represented demand changes to its new block power level. More generally, integrating $\Delta P^{\mathrm{blk}}(t)$ over the $[\tau_{n-1}, \tau_n]$ gives the net energy imbalance, whereas integrating its absolute value gives the absolute energy imbalance.

The example shows that power balance at the market interval endpoints does not guarantee intra-interval power balance, even when the demand transition is known in advance. A disconnection-related transition can be illustrated analogously. This motivates retaining the temporal evolution of planned demand transitions and coordinating generation with the declared demand-side power trajectory.

### *B. Workload-Driven Demand Variations*

Beyond planned connection and disconnection transitions, workload execution can produce data center demand variations across different timescales [15]. The upper panels of Fig. 2 show representative workload power traces derived from [16]. In these examples, training and fine-tuning exhibit sustained operating periods with repeated power variations, whereas inference exhibits short and intermittent peaks. These traces illustrate workload-dependent temporal characteristics. In Fig. 2, the background demand and its matching generation are omitted from these schematics.

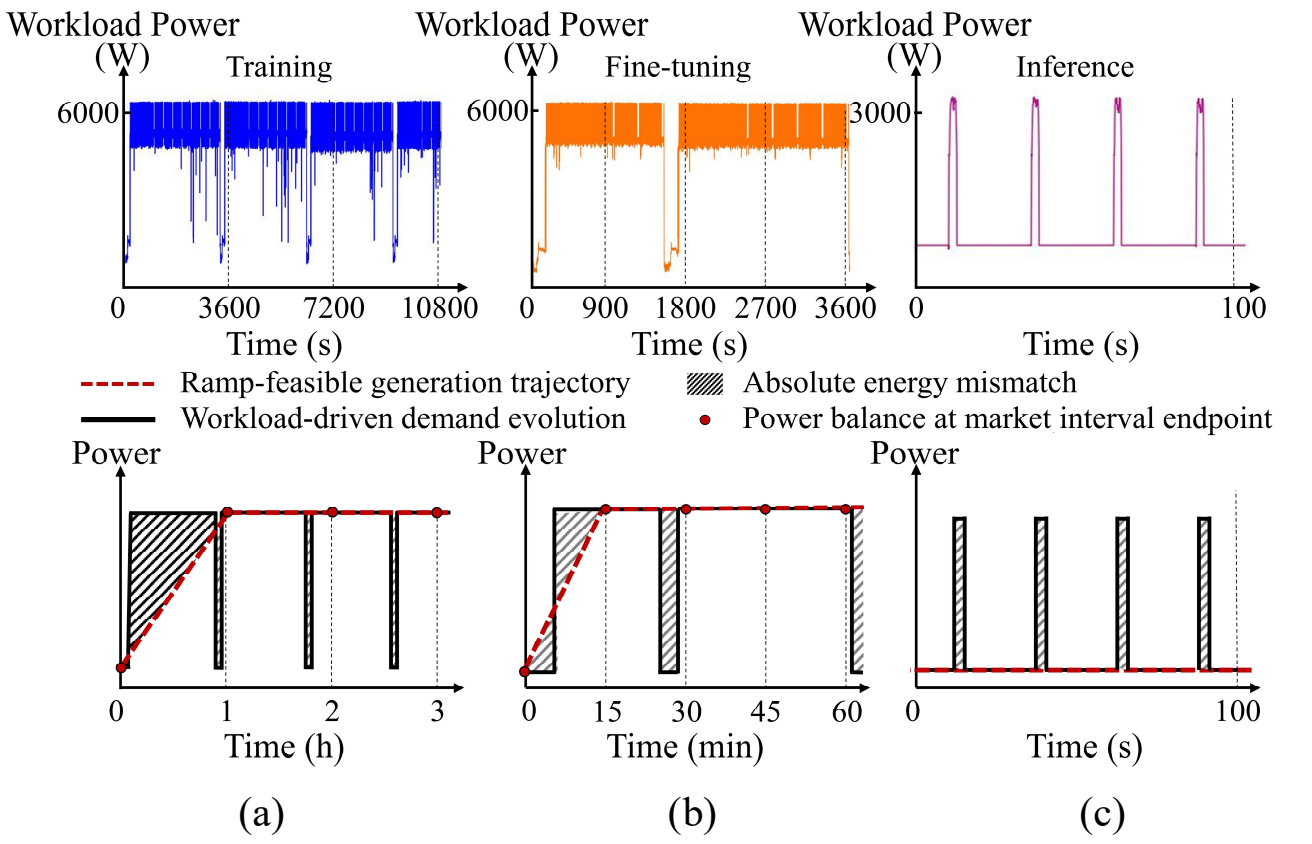

Fig. 2. Workload-driven power variations and schematic generation-demand mismatches: (a) training; (b) fine-tuning; and (c) inference. Upper panels show workload power traces derived from [16]. Lower panels illustrate mismatches between endpoint-based generation reconstructions and underlying demand evolution in (a) and (b), and residual short-timescale deviations excluded from the day-ahead declaration in (c).

For day-ahead market participation, the relevant criterion is whether aggregate demand evolution can be specified before market clearing. Workload schedules and planned computing activities may provide information for constructing such a declaration. The proposed framework therefore focuses on predictable aggregate demand evolution, while deviations between realized and declared demand remain outside the deterministic day-ahead clearing model.

For illustration, the lower panels of Fig. 2(a) and Fig. 2(b) schematically show how interval-based demand representations can omit intra-interval demand evolution. Fig. 1 compares generation with the block demand representation, whereas Fig. 2 compares endpoint-based generation reconstructions with the underlying workload-driven demand. The workload-driven demand evolution is shown together with a piecewise-linear ramp-feasible generation reconstruction that matches the retained endpoint power values without necessarily matching the underlying demand throughout each interval.

These two panels illustrate how the representation issue can arise at different temporal resolutions. In Fig. 2(a), an hourly interval-based representation captures the demand level assigned to each hour but does not preserve its intra-hour evolution. The resulting generation trajectory matches the represented demand at the hourly endpoints but differs from the underlying demand evolution within the interval, producing the shaded intra-interval energy imbalance. Fig. 2(b) illustrates the same mechanism using shorter 15-min market intervals. The shaded areas measure the absolute energy mismatch between the illustrated generation and the workload-driven demand.

Fig. 2(c) provides a short-timescale contrast. The intermittent inference peaks occur on a much faster timescale and are not assumed here to be specified as part of the day-ahead declaration. They are therefore treated as residual short-timescale deviations between realized and declared demand and are left to subsequent balancing mechanisms.

### *C. Demand-Side Power Trajectories for Data Centers*

The declared demand-side power trajectory, denoted by $L(t)$, specifies the power demand of a data center used in day-ahead market clearing. Unlike an interval energy quantity alone, $L(t)$ specifies the power level throughout the market-clearing horizon. It is constructed from information available before clearing and retains predictable aggregate demand evolution.

From the perspective of the market participant, the data center constructs $L(t)$ based on its workload schedules, planned tasks, and expected power consumption[1]. Here, the declared trajectory is required to remain continuous. In the case studies, $L(t)$ is represented by a continuous, piecewise-polynomial function whose polynomial expression and slope may change across segments while its value remains continuous at segment endpoints. This representation accommodates the planned continuous transitions in Section II-A and the predictable aggregate demand variations in Section II-B.

From the perspective of the system operator, $L(t)$ is an exogenous demand input rather than a decision variable. Continuity alone does not guarantee feasibility: the total demand trajectory $D(t)+L(t)$ must admit generation trajectories satisfying the modeled constraints. Deviations between realized data center demand and $L(t)$ are outside the deterministic day-ahead clearing model and are left to subsequent balancing mechanisms. Section III constructs the generation trajectories and the corresponding price trajectory for feasible clearing outcomes, while Section IV evaluates the settlement effects.

## III. Trajectory-Based Energy Market Clearing

This section incorporates the declared demand-side power trajectory into day-ahead energy clearing as an exogenous input. The trajectory-construction methodology in [13] and its dual extension in [14] are adopted to obtain generation trajectories and the corresponding price trajectory. To focus on the effects of temporal representation, unit commitment decisions are treated as fixed, while network constraints, reserve procurement, and other market processes are excluded.

### *A. Continuous-time Formulation*

The continuous-time formulation in [13] is adopted with the total demand trajectory $D(t)+L(t)$. Here, $D(t)$ denotes the baseline system demand excluding the data center demand, and

[1]Auxiliary loads, including those associated with cooling and data-storage equipment, are treated as exogenous background contributions within $L(t)$. The workload power traces in Fig. 2 illustrate computing-workload power variations rather than complete facility demand.

$L(t)$ denotes the declared demand-side power trajectory of the data center. Both trajectories are treated as given in advance, and the system is modeled on an aggregate basis without network differentiation.

Let $\mathcal{T}=[S,T]$ denote the market-clearing horizon, with start and end instants $S$ and $T$, respectively. Given the generation output levels prescribed at these boundaries, the objective is to minimize total generation cost while satisfying power balance and generation operating constraints over $\mathcal{T}$. The continuous-time dispatch problem is formulated as follows:

$$\min_{\mathbf{G}(t)} \int_{\mathcal{T}} \left( C\left(\mathbf{G}(t)\right)\right) dt \tag{2}$$

$$\text{s.t. } \sum_{k=1}^{K} G_k(t) = D(t) + L(t), \forall t \in \mathcal{T}, \tag{3}$$

$$G_k(t) \in \mathcal{G}_k, k = 1,...,K, \forall t \in \mathcal{T}, \tag{4}$$

$$\mathcal{G}_k = \left\{ G_k(t) \;\middle|\; \begin{array}{c} G_k^{\min} \le G_k(t) \le G_k^{\max} \\ R_k^{\text{down}} \le G_k'(t) \le R_k^{\text{up}} \\ G_k(S) = G_{k,S}, G_k(T) = G_{k,T} \end{array} \right\}. \tag{5}$$

where $\mathbf{G}(t)=[G_1(t), ..., G_K(t)]^\top$ denotes the vector of generation trajectories, and $G_k(t)$ denotes the trajectory of unit $k$. The aggregate generation cost rate is $C(\mathbf{G}(t))=\Sigma_{k=1}^{K} C_k(G_k(t))$, where $C_k(\cdot)$ is the generation cost rate of unit $k$. Integrating this rate over $\mathcal{T}$ gives the total generation cost in (2). $G_k^{\min}$ and $G_k^{\max}$ are the minimum and maximum generation capacities. $R_k^{\text{down}}$ and $R_k^{\text{up}}$ are the lower and upper bounds on the ramping rate, measured in MW per unit of time. The derivative $G_k'(t)$ represents the ramping rate of unit $k$. The set $\mathcal{G}_k$ is the generation operating constraints of unit $k$ over $\mathcal{T}$. $G_{k,S}$ and $G_{k,T}$ are the given generation output levels at $t=S$ and $t=T$.

$D(t)+L(t)$ may be represented as a continuous piecewise-defined function. Accordingly, the ramping constraints in (5) are imposed almost everywhere, allowing the generation trajectories to change slope at segment endpoints while preserving continuity of the generation output. For the linear generation cost rates used in the settlement analysis, the dispatch problem can be written in the compact operator form in (6), following the primal representation in [14]. The primal and dual variables used for trajectory construction are introduced in Section III-B.

$$\begin{aligned} \min_{x(t)} & \int_{\mathcal{T}} f^\top x(t) dt \\ \text{s.t.} \quad & A_{\text{eq}} x(t) = b_{\text{eq}} \\ & A_{\text{ie}} x(t) \ge b_{\text{ie}} \end{aligned} \tag{6}$$

where $x(t)$ collects the generation trajectories, and $f$ contains the associated bid or marginal-cost coefficients. $A_{\text{eq}}$ and $A_{\text{ie}}$ denote the linear equality and inequality operators, respectively, while $b_{\text{eq}}$ and $b_{\text{ie}}$ denote the prescribed right-hand-side data; in particular, the power-balance component of $b_{\text{eq}}$ is $D(t)+L(t)$.

## *B. Iterative Dispatch Methodology*

The compact form in (6) remains infinite-dimensional because its decision variables are functions of time. To construct computationally tractable trajectory representations, the iterative dispatch methodology in [13] and its dual extension in [14] are adopted. The procedure alternates between local parametric trajectory construction and constraint verification.

A two-dimensional parameter vector $\theta=[t, D^P]^\top$ is introduced, where $t$ is the time coordinate and $D^P$ is an independent coordinate representing aggregate demand. Treating time and demand as separate parameters preserves the affine parameter dependence of the local constraint right-hand sides. The prescribed total demand trajectory $D(t)+L(t)$ defines the following path in this parameter space:

$$\theta^C(t) = [t, D(t) + L(t)]^\top, \forall t \in \mathcal{T} \tag{7}$$

where the superscript $C$ denotes quantities associated with the trajectory-based clearing. The total demand path $\theta^C(t)$ remains fixed during the iterations; only the local formulations and their parametric solution maps are updated.

Let $\mathcal{T}^U=[S^U, T^U]\subseteq\mathcal{T}$ denote a local updating range, which is restricted to the portion of the time horizon requiring update based on the latest constraint-verification stage. The generation power levels at $S^U$ and $T^U$ are prescribed for each local trajectory-construction problem, using initialization data or the results of the latest discrete-time adaptive dispatch [13].

Over $\mathcal{T}^U$, the trajectory-construction stage uses the finite-dimensional parametric linear program and its dual in (8)[2], following [14]. The local problem data, including the bid or marginal-cost coefficients $f^U$ and the prescribed endpoint output levels, are fixed during each local construction step.

$$\begin{aligned} &\text{(Primal)} && \min_{x(\theta)} (f^U)^\top x(\theta) \\ &\text{s.t.} && \lambda(\theta): A_{\text{eq}} x(\theta) = b_{\text{eq}} + E_{\text{eq}}\theta \\ & && \mu(\theta): A_{\text{ie}} x(\theta) \ge b_{\text{ie}} + E_{\text{ie}}\theta \\ &\text{(Dual)} && \max_{\lambda(\theta),\mu(\theta)} \left(b_{\text{eq}} + E_{\text{eq}}\theta\right)^\top \lambda(\theta) + \left(b_{\text{ie}} + E_{\text{ie}}\theta\right)^\top \mu(\theta) \\ &\text{s.t.} && A_{\text{eq}}^\top \lambda(\theta) + A_{\text{ie}}^\top \mu(\theta) = f^U, \mu(\theta) \ge 0, \end{aligned} \tag{8}$$

where $x(\theta)$, $\lambda(\theta)$, and $\mu(\theta)$ are the parameter-dependent primal and dual vectors. Unlike the operators in (6), $A_{\text{eq}}$ and $A_{\text{ie}}$ in (8) are finite-dimensional constraint matrices. The matrices $E_{\text{eq}}$ and $E_{\text{ie}}$ describe the affine dependence of the right-hand sides on $\theta$. The remaining right-hand-side terms $b_{\text{eq}}$ and $b_{\text{ie}}$ collect the parameter-independent components. The local-model index is suppressed for notational simplicity.

In the local model, $D^P$ enters the power-balance equation, while the prescribed endpoint outputs and other local data are held fixed. Following [13], ramping is represented through reachability constraints linking the current output to the outputs at $S^U$ and $T^U$. These constraints do not replace the subsequent verification of the continuous-time derivative bounds in (5).

Solving the local parametric linear problem identifies critical regions and their associated primal-dual solution maps. Intersecting these regions with the prescribed path $\theta^C(t)$ determines the trajectory segments within the updating range. Evaluating the corresponding solution maps along the path $\theta^C(t)$ yields candidate generation and price trajectories.

In the constraint-verification stage, the derivatives of the candidate generation trajectories are checked against the ramping rates in (5), and continuity is independently checked at segment endpoints. A discrete-time adaptive dispatch model determines revised endpoint output levels and the updating range for the next iteration [13]. Previously verified portions outside the updating range are retained. The construction and verification stages are repeated until the feasibility checks and the prescribed endpoint-mismatch tolerance are satisfied; the detailed verification and updating rules are given in [13], [14].

[2]Trajectory construction and pricing use the finite-dimensional local primal-dual problems in (8), with continuous-time feasibility checked separately. No equivalence between the assembled trajectories and a globally optimal primal-dual solution of the original infinite-dimensional problem is asserted.

For each retained trajectory segment, let $x^*(\theta)$ and $\lambda^*(\theta)$ denote selected optimal primal and equality-constraint dual solutions of the corresponding local problem in (8). With the local-model and critical-region indices suppressed, the time-indexed outcomes are obtained as follows:

$$x^C(t) = x^*\left(\theta^C(t)\right), \lambda^C(t) = \lambda^*\left(\theta^C(t)\right). \quad (9)$$

Here, $x^C(t)$ defines the resulting generation trajectories, and $\lambda^C(t)$ is adopted as the system-wide marginal energy price under the local parametric formulation in [14]. $\lambda^C(t)$ has units of dollars per MW per unit of time. This price interpretation is conditional on the endpoint output levels and other data of the local construction problem. With time and all other local data fixed, the selected power-balance multiplier is a shadow price of the local optimal cost rate with respect to $D^P$. As in [13], [14], the iterative dispatch methodology does not guarantee global optimality of the original continuous-time dispatch problem.

A feasible trajectory may not exist if the prescribed demand trajectory and boundary output levels are incompatible with the modeled generation operating constraints. When the feasibility checks and prescribed endpoint-mismatch tolerance are satisfied, (9) provides piecewise-defined generation and price trajectories over the market-clearing horizon, rather than values only at isolated dispatch instants.

### C. Trajectory Properties Relevant to Settlement

The retained local parametric solution maps, evaluated along the prescribed demand path, provide a finite piecewise representation of the market-clearing outcomes. For settlement, a common temporal partition is formed from critical-region transition instants, segment endpoints, and any breakpoints in the prescribed demand representation. Let $S=t_0<t_1<\cdots<t_N=T$ denote the resulting partition endpoints, and define trajectory intervals as $\mathcal{T}_n=[t_n, t_{n+1}]$, $n=0, ..., N-1$. Within the interior of each trajectory interval, the applicable local problem data and selected critical-region solution map remain unchanged.

The continuous-time formulation requires generation trajectories to remain continuous across adjacent segments. In the numerical implementation, the endpoint power levels of adjacent segments are checked for agreement within the prescribed endpoint-mismatch tolerance using the procedure in [13]. The price trajectory need not be continuous and may jump when the applicable local solution map changes [14]. Under the adopted representation, such price jumps are confined to the common partition endpoints. With network constraints omitted, the selected power-balance multiplier defines the system-wide marginal energy price trajectory $\lambda^C(t)$.

For the piecewise-defined price trajectory, the values assigned at the finitely many partition endpoints do not affect the settlement integrals. For definiteness, the left-hand limit is used at partition endpoints after $S$, consistent with [14]. Energy quantities are evaluated by integrating the cleared demand and generation trajectories over each $\mathcal{T}_n$, while energy payments and energy revenues are evaluated by integrating their products with the associated price trajectory over the same $\mathcal{T}_n$. The resulting interval quantities are then summed over $\mathcal{T}$.

## IV. Trajectory-based Settlement Analysis

The trajectory intervals defined in Section III-C provide a common temporal partition over the market-clearing horizon. Building on Section III, this section defines physical and economic temporal consistency. It then examines three distinct aspects of settlement analysis: energy payments and generator revenues, supply-side compensation, and the monetary magnitude of representation-induced power imbalance. This separation helps interpret payment changes without equating them with changes in generation cost. The analysis extends the payment distinctions in [17] to continuous-time accounting.

### A. Physical and Economic Temporal Consistency

Interval energy quantities or endpoint power levels alone do not uniquely specify intra-interval demand evolution. Matching interval energy quantities or balancing supply and demand only at market interval endpoints therefore does not, by itself, establish power balance and generation feasibility throughout the interval. Moreover, using different temporal representations for dispatch and prices may alter the time-matched pairing of cleared power and its associated price in settlement.

To distinguish these physical and settlement effects, two notions of temporal consistency are introduced within the power-trajectory framework. These notions concern temporal representation and dispatch-price correspondence, rather than numerical solution accuracy.

*Definition 1 Physical Temporal Consistency*: A market-clearing outcome satisfies physical temporal consistency if its generation trajectories balance the total demand trajectory used in market clearing throughout $\mathcal{T}$ and satisfy all modeled generation operating constraints. In this paper, these constraints include generation output limits, ramping constraints, and the prescribed horizon boundary conditions.

*Definition 2 Economic Temporal Consistency*: A settlement outcome satisfies economic temporal consistency if the dispatch-price correspondence established by the adopted clearing and pricing formulation is preserved in energy settlement on the common temporal basis. For trajectory-based outcomes, this requires pairing cleared power with its associated marginal energy price at each instant and integrating their product over $\mathcal{T}$ to evaluate energy payments.

Physical temporal consistency concerns continuous-time power balance and satisfaction of the modeled generation operating constraints, whereas economic temporal consistency concerns preservation of the dispatch-price correspondence in energy settlement. The interval-based demand representations discussed in Section II illustrate the physical inconsistency issue. A separate settlement inconsistency can arise when a physically feasible dispatch is settled using an interval-based price representation that does not preserve the marginal prices associated with its cleared power. Refining the temporal partition alone does not guarantee the required dispatch-price correspondence, and a difference in temporal partitioning alone does not establish inconsistency. Moreover, economic temporal consistency does not, by itself, ensure cost recovery or eliminate the supply-side compensation defined below.

### B. Energy Quantities, Prices, and In-Market Energy Payment

Consider a feasible market-clearing outcome in which the prescribed, inelastic total demand trajectory $D(t)+L(t)$ is served in full. The energy quantities and payments defined below refer to this cleared total demand trajectory $D(t)+L(t)$ rather than realized consumption.

*Cleared energy quantities:* An interval energy quantity specifies the amount of energy assigned to a market interval but does not, by itself, specify the intra-interval power profile. In the power-trajectory framework, $D(t)+L(t)$ is retained, and the cleared demand energy over each trajectory interval $\mathcal{T}_n$ is obtained by integration. The interval and horizon-wide energy quantities are defined as follows:

$$E_{\mathcal{T}}^{C}=\sum_{n=0}^{N-1}E_n^C,\ E_n^C=\int_{\mathcal{T}_n}\big(D(t)+L(t)\big)dt. \tag{10}$$

Here, $E_n^C$ is the aggregate cleared demand energy over $\mathcal{T}_n$, and $E_{\mathcal{T}}^C$ is the total cleared demand energy over $\mathcal{T}$.

*Price trajectory*: The energy price used for settlement is $\lambda^C(t)$, $t\in\mathcal{T}$, as defined in Section III-B. Unlike a single price assigned to a market interval, $\lambda^C(t)$ can represent price changes within that interval. It is evaluated together with $D(t)+L(t)$ on the common temporal partition defined in Section III-C.

*In-market Energy Payment*: Over $\mathcal{T}$, the aggregate demand-side energy payment is obtained by integrating the product of cleared demand and its price trajectory at the same time:

$$\Pi_{\mathcal{T}}^{\text{EP}}=\sum_{n=0}^{N-1}\int_{\mathcal{T}_n}\lambda^C(t)\big(D(t)+L(t)\big)dt. \tag{11}$$

$\Pi_{\mathcal{T}}^{\text{EP}}$ represents the aggregate in-market energy payment for all cleared demand. It excludes the supply-side compensation evaluated separately later.

With power measured in MW and time in hours, the cleared energy quantities in (10) are measured in MWh, $\lambda^C(t)$ in dollars per MWh, and $\Pi_{\mathcal{T}}^{\text{EP}}$ in dollars. For fixed demand and price trajectories, further subdivision of the common temporal partition changes neither the total cleared demand energy nor the in-market energy payment. For the feasible outcome considered here, the total demand trajectory remains continuous but may be piecewise-defined, whereas the price trajectory may jump at finitely many partition endpoints. As discussed in Section III-C, the values assigned at these isolated endpoints do not affect the settlement integrals.

### C. Self-Scheduling and Supply-Side Compensation

The analysis distinguishes positive-margin output-increase opportunities from gross below-cost production. To evaluate the former, unit $k$ solves a price-taking self-scheduling problem[3] at the given price trajectory $\lambda^C(t)$. The unit retains the constraints in $\mathcal{G}_k$, with its output at every internal partition endpoint fixed to the cleared value. Under these constraints, the problem maximizes the unit's energy-market profit over $\mathcal{T}$:

$$\max_{x_k^S(t)}\sum_{n=0}^{N-1}\int_{\mathcal{T}_n}\Big(x_k^S(t)\big(\lambda^C(t)-f_k\big)\Big)dt$$

$$\text{s.t.}\qquad x_k^S(t)\in\mathcal{X}_k,\forall t\in\mathcal{T},$$

$$\mathcal{X}_k=\Big\{x(t)\in\mathcal{G}_k\,\Big|\,x(t_n)=x_k^C(t_n),n=1,\ldots,N-1\Big\}. \tag{12}$$

Here, $x_k^S(t)$ denotes a selected optimal self-scheduling trajectory, and $x_k^C(t)$ denotes the cleared generation trajectory obtained in Section III. The generation cost rate is taken to be linear with a constant coefficient $f_k$[4]. The market clearing horizon boundary conditions at $t_0=S$ and $t_N=T$ are inherited from $\mathcal{G}_k$. This benchmark does not impose system-wide power balance or recompute prices in response to the unit's deviation.

Since $x_k^S(t)\geqslant G_k^{\min}\geqslant 0$, a negative energy margin does not imply negative generation. Instead, the unit seeks to reduce its output subject to $\mathcal{X}_k$. The unit-level settlement quantities are defined by $x_k^C(t)$, $\lambda^C(t)$ and $f_k$ with $x_k^S(t)$ additionally used in (14).

*Energy Revenue* (ER): The in-market energy revenue of unit $k$ is obtained by integrating the product of its cleared output and the associated price trajectory. This revenue is measured before deducting the modeled production costs.

$$\Upsilon_{k,\mathcal{T}}^{\text{ER}}=\sum_{n=0}^{N-1}\int_{\mathcal{T}_n}\lambda^C(t)x_k^C(t)dt. \tag{13}$$

*Positive-Margin Lost Opportunity Cost* (PMLOC): PMLOC measures the positive-margin output-increase contribution identified in the selected optimal self-scheduling trajectory relative to the cleared generation trajectory:

$$\Upsilon_{k,\mathcal{T}}^{\text{PMLOC}}=\sum_{n=0}^{N-1}\int_{\mathcal{T}_n}\big(\lambda^C(t)-f_k\big)^+\big(x_k^S(t)-x_k^C(t)\big)^+dt, \tag{14}$$

where $(x)^+=\max(x,0)$. $\Upsilon_{k,\mathcal{T}}^{\text{PMLOC}}=0$ indicates that the selected $x_k^S(t)$ identifies no positive-margin output-increase contribution; it does not rule out other profitable deviations.

*Extra-Marginal Compensation* (EMC): EMC is defined as the gross production-cost shortfall associated with cleared output when the price is below the modeled marginal cost:

$$\Upsilon_{k,\mathcal{T}}^{\text{EMC}}=\sum_{n=0}^{N-1}\int_{\mathcal{T}_n}\big(f_k-\lambda^C(t)\big)^+x_k^C(t)dt. \tag{15}$$

The positive part is taken before integration, so positive energy margins at other times do not offset these shortfalls.

Under the definition in [19], lost opportunity cost (LOC) compares maximum self-scheduling profit with cleared-dispatch profit at the given prices. LOC can be calculated as

$$LOC_k=\max_{x_k^S(t)\in\mathcal{X}_k}\sum_{n=0}^{N-1}\int_{\mathcal{T}_n}\big(\lambda^C(t)-f_k\big)\big(x_k^S(t)-x_k^C(t)\big)dt \tag{16}$$

For the benchmark in (12), positive local contributions to the profit difference arise from two cases: increasing output at a positive energy margin and reducing output at a negative energy margin. PMLOC measures the first type of deviation. EMC includes the shortfall avoided by the second type of deviation and additionally accounts for below-cost production common to the cleared and selected self-scheduling trajectories.

This component-wise accounting separates positive-margin opportunities from below-cost production before aggregation over $\mathcal{T}$. Evaluating the integrands along the generation and price trajectories identifies when these contributions arise and supports their attribution to minimum generation limits, ramping constraints, and price transitions. The resulting analysis complements, rather than replaces, net-profit analysis.

[3] Conventional self-scheduling problems are formulated over discrete market intervals or time steps, as in [8] and [18]. For continuous-time results here, cleared endpoint outputs are fixed to evaluate intra-segment deviations under the given price trajectory. The benchmark does not assess deviations that change the internal endpoint outputs and is not intended as a complete analysis of generator incentives or strategic behavior.

[4] Submitted bids are assumed to equal the corresponding marginal production costs, and strategic bidding is outside the scope of this paper. Accordingly, $f_k$ denotes the fixed marginal-cost coefficient associated with the generation-output term of unit $k$ in the linear cost representation here.

*Generator Revenue* (GR): Under the compensation convention specified in Section IV-D, the PMLOC and EMC are assumed to be paid in full in addition to energy revenue.

$$\Upsilon^{\mathrm{GR}}_{k,\mathcal{T}} = \Upsilon^{\mathrm{ER}}_{k,\mathcal{T}} + \Upsilon^{\mathrm{PMLOC}}_{k,\mathcal{T}} + \Upsilon^{\mathrm{EMC}}_{k,\mathcal{T}}. \tag{17}$$

## D. Customer Payments and Monetary Mismatch

For the settlement comparisons, PMLOC and EMC are assumed to be paid in full in addition to energy revenues and recovered from aggregate demand-side charges. This accounting convention is not a payment requirement implied by the clearing model. The system-wide payments are:

$$\Pi^{\mathrm{PMLOC}}_{\mathcal{T}} = \sum\nolimits_{k=1}^{K} \Upsilon^{\mathrm{PMLOC}}_{k,\mathcal{T}}, \Pi^{\mathrm{EMC}}_{\mathcal{T}} = \sum\nolimits_{k=1}^{K} \Upsilon^{\mathrm{EMC}}_{k,\mathcal{T}}. \tag{18}$$

*Total Customer Payment*: Under the assumed convention, the total modeled charge assigned to demand-side participants is:

$$\Pi_{\mathcal{T}} = \Pi^{\mathrm{EP}}_{\mathcal{T}} + \Pi^{\mathrm{OMP}}_{\mathcal{T}}, \Pi^{\mathrm{OMP}}_{\mathcal{T}} = \Pi^{\mathrm{PMLOC}}_{\mathcal{T}} + \Pi^{\mathrm{EMC}}_{\mathcal{T}}. \tag{19}$$

Here, the out-of-market payment[5] $\Pi^{\mathrm{OMP}}_{\mathcal{T}}$ comprises the assumed PMLOC and EMC transfers, while $\Pi^{\mathrm{EP}}_{\mathcal{T}}$ is the in-market energy payment defined in (11).

When aggregate generation balances the represented demand throughout $\mathcal{T}$ and the same time-indexed settlement price is applied to both sides, aggregate in-market energy payments equal aggregate energy revenues. In the comparison scenarios, the energy-settlement residual is aggregate energy revenue minus aggregate in-market energy payment. Positive and negative monetary contributions can offset in this residual. The following diagnostic measure quantifies the gross monetary magnitude of representation-induced power imbalance between ramp-feasible generation and represented demand.

*Representation-induced Monetary Mismatch* (RMM): Let $s$ index the comparison scenarios, and let $d_s(t)$, $G_{k,s}(t)$ and $\lambda_s(t)$ denote the demand representation, generation trajectory of unit $k$, and the settlement price used in scenario $s$ respectively. The representation-induced power imbalance is defined as:

$$\Delta P^{\mathrm{rep}}_{s}(t) = \sum\nolimits_{k=1}^{K} G_{k,s}(t) - d_s(t), \forall t \in \mathcal{T}. \tag{20}$$

Using the settlement price assigned to each scenario, RMM is evaluated as follows:

$$\Pi^{\mathrm{RMM}}_{\mathcal{T},s} = \int_{\mathcal{T}} \left| \lambda_s(t) \Delta P^{\mathrm{rep}}_{s}(t) \right| dt. \tag{21}$$

For scenarios using the interval-based demand representation, $d_s(t)=D(t)+L^{\mathrm{blk}}(t)$, and $\Delta P^{\mathrm{rep}}_{s}(t)$ has the form defined in (1). Taking the absolute value before integration prevents positive and negative contributions from offsetting. RMM therefore measures their gross monetary magnitude; it is neither a signed settlement surplus or deficit nor the cost of balancing actions.

The mismatch trajectory in (20) identifies the temporal evolution of the power imbalance, while RMM summarizes its gross monetary magnitude in (21). RMM is reported separately as a diagnostic measure and is not included in the total customer payment defined in (19). The financing or allocation of any signed energy-settlement residual in the comparison scenarios is not modeled. If generation balances the represented demand throughout $\mathcal{T}$, RMM is zero; however, this does not imply that the separately defined PMLOC and EMC components vanish. The diagnostic value of this separation is to distinguish representation-induced monetary mismatch from below-cost production under the adopted dispatch and pricing formulation. It identifies the units and time intervals contributing to each component and helps avoid interpreting changes in compensation payments as changes in production efficiency.

[5]The out-of-market payment refers to the assumed PMLOC and EMC transfers in addition to in-market energy payments here. This convention is not intended to reproduce the uplift rules of any specific market.

# V. Case Studies

The power-trajectory framework is examined using an illustrative 5-bus system and a modified IEEE RTS-GMLC system. The 5-bus case compares four dispatch-price pairings based on interval-based and trajectory-based representations for a planned connection-related data center demand transition. This case examines the settlement effects of different temporal representations and illustrates the minimum-output and ramp-related components of EMC. The modified RTS-GMLC case extends the analysis to a larger set of generating units under a constructed workload-driven data center demand trajectory.

For both systems, the system parameters, continuous-time load profiles, generation and price trajectories, settlement quantities, and self-scheduling results are provided in [20]. Settlement integrals are evaluated with time expressed in hours, whereas the figures display time in minutes.

## A. Results in the Illustrative 5-bus System

The illustrative 5-bus case considers a one-hour market-clearing horizon with demand treated as an exogenous input. The case examines generation trajectories under different demand representations and the settlement effects of the selected dispatch-price pairings.

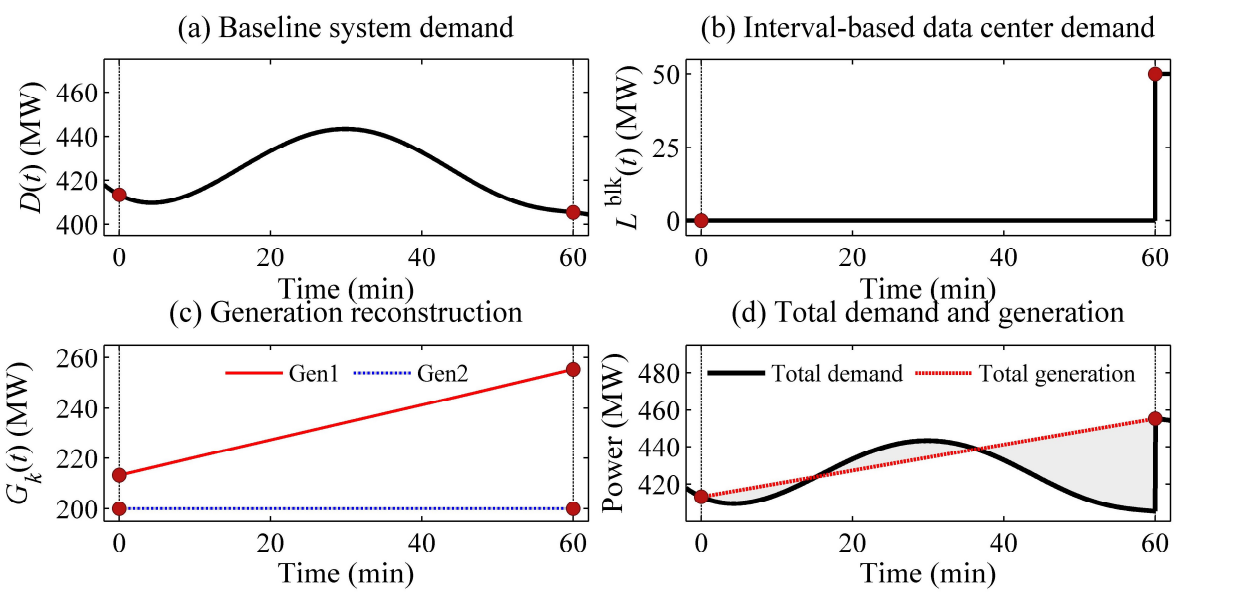

Fig. 3. Endpoint-based benchmark for a planned connection-related demand transition: (a) baseline system demand; (b) data center demand; (c) reconstructed generation trajectories; and (d) total demand and generation.

Fig. 3 presents an endpoint-based benchmark using an interval-based demand representation for the data center. The baseline system demand $D(t)$ is shown in Fig. 3(a). In Fig. 3(b), the data center demand is represented by the piecewise-constant load profile $L^{\mathrm{blk}}(t)$, with a step change at $t$=60 min. The benchmark determines unit-level generation outputs at $t$=0 and $t$=60 min, as indicated by the red markers in Fig. 3(c). The connecting curves in Fig. 3(c) are selected linear ramp-feasible reconstructions between these cleared endpoint outputs. Although aggregate generation matches the represented total demand at both endpoints, the two differ within this hour, as shown in Fig. 3(d). Subtracting the represented total demand from aggregate generation gives the representation-induced power imbalance defined in (1). Its gross monetary magnitude is evaluated using RMM in the settlement comparison below.

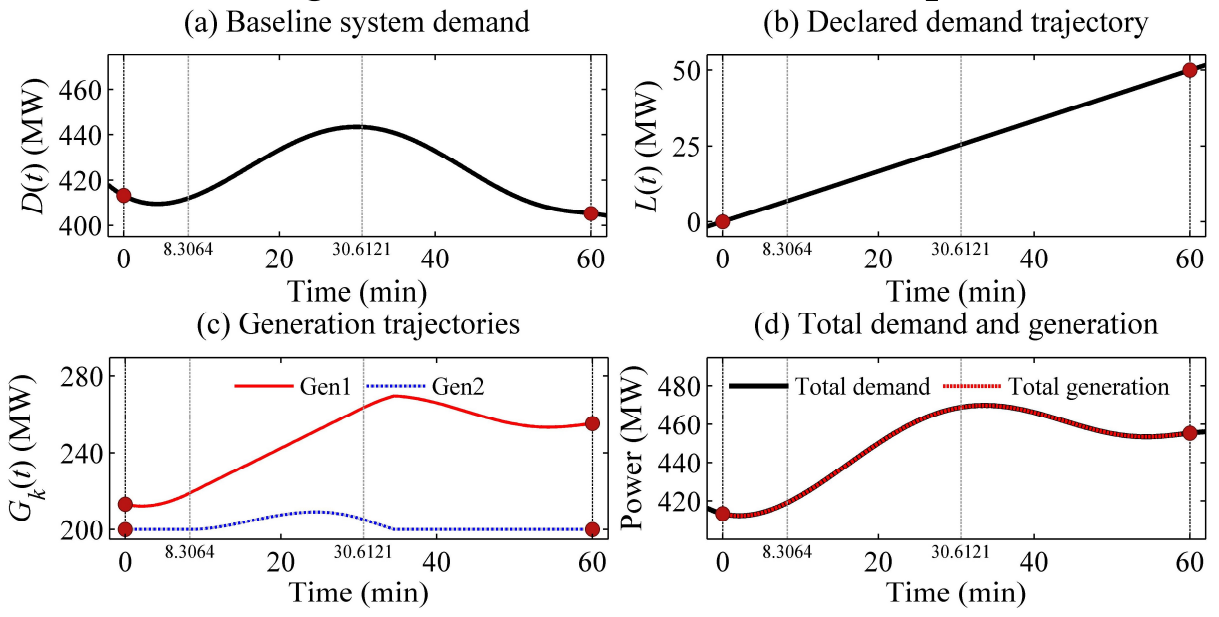

Fig. 4. Trajectory-based clearing for a planned connection-related demand transition: (a) baseline system demand; (b) declared demand-side power trajectory; (c) generation trajectories; and (d) total demand and generation.

In the power-trajectory framework, data center demand is specified by the declared demand-side power trajectory $L(t)$ in Fig. 4(b). A linear trajectory is used in this example, but the formulation is not restricted to linear demand trajectories. The generation trajectories in Fig. 4(c) are constructed to balance $D(t)+L(t)$ subject to the modeled generation operating constraints. Fig. 4(d) shows their aggregate output matching the prescribed total demand trajectory throughout the market-clearing horizon. The resulting market-clearing outcome has no representation-induced power imbalance relative to $D(t)+L(t)$ and satisfies physical temporal consistency under Definition 1.

TABLE I
FOUR DISPATCH-PRICE PAIRINGS FOR SETTLEMENT ANALYSIS

| Scenario | Total Demand | Generation trajectories | Settlement Price |
|---|---|---|---|
| S1 | $D(t)+L^{blk}(t)$ | Endpoint-based reconstruction in Fig. 3(c) | Interval-based price in Fig. 5(e) |
| S2 | $D(t)+L(t)$ | Trajectory-based dispatch in Fig. 4(c) | Interval-based price in Fig. 5(e) |
| S3 | $D(t)+L^{blk}(t)$ | Endpoint-based reconstruction in Fig. 3(c) | Trajectory-based price in Fig. 5(f) |
| S4 | $D(t)+L(t)$ | Trajectory-based dispatch in Fig. 4(c) | Trajectory-based price in Fig. 5(f) |

Table I summarizes four dispatch-price pairings used to examine the settlement effects of temporal representation. The price trajectory in Fig. 5(f) is associated with the dispatch in Fig. 4(c); S4 therefore preserves the dispatch-price correspondence. S2 and S3 are counterfactual settlement pairings rather than independently cleared market outcomes. The interval-based price is obtained from the endpoint-based benchmark rather than by averaging the trajectory-based price. The S1-S3 and S2-S4 comparisons change the settlement price while holding demand and generation trajectories fixed. All scenarios use the common temporal partition and the accounting definitions in Section IV, with self-scheduling endpoints fixed to the corresponding dispatch values. The demand representations in Fig. 3(b) and Fig. 4(b) agree at $t$=0 and $t$=60 min but are not energy-equivalent over the hour. Comparisons across the two demand-dispatch representations therefore do not represent alternatives serving the same hourly demand energy.

Tables II and III report the generator-side settlement quantities and the system-wide settlement outcomes, respectively. Payment differences are calculated from unrounded values. The marginal-cost coefficients of Gen1 and Gen2 are \$25/MWh and \$30/MWh, respectively. For the selected reconstruction, S1 and S3 have the same absolute intra-interval energy imbalance of 14.7416 MWh relative to $D(t)+L^{blk}(t)$. S2 and S4 balance the prescribed total demand trajectory $D(t)+L(t)$, with zero absolute energy imbalance and RMM within numerical tolerance. Together with the numerical verification of the modeled generation operating constraints within the prescribed tolerance, this establishes physical temporal consistency for S2 and S4 under Definition 1.

TABLE II
GENERATOR-SIDE SETTLEMENT COMPONENTS UNDER DIFFERENT TEMPORAL REPRESENTATIONS

| Unit | Quantity($) | S1 | S2 | S3 | S4 |
|---|---|---|---|---|---|
| Gen1 | $\Upsilon_{k,\mathcal{T}}^{ER}$ | 5.8576e03 | 6.1749e03 | 6.2794e03 | 6.6235e03 |
| | $\Upsilon_{k,\mathcal{T}}^{PMLOC}$ | 0 | 0 | 1.8186e01 | 0 |
| | $\Upsilon_{k,\mathcal{T}}^{EMC}$ | 0 | 0 | 0 | 0 |
| | $\Upsilon_{k,\mathcal{T}}^{GR}$ | 5.8576e03 | 6.1749e03 | 6.2976e03 | 6.6235e03 |
| Gen2 | $\Upsilon_{k,\mathcal{T}}^{ER}$ | 5.0000e03 | 5.0562e03 | 5.3718e03 | 5.4383e03 |
| | $\Upsilon_{k,\mathcal{T}}^{PMLOC}$ | 0 | 0 | 0 | 0 |
| | $\Upsilon_{k,\mathcal{T}}^{EMC}$ | 1.0000e03 | 1.0112e03 | 6.2824e02 | 6.2909e02 |
| | $\Upsilon_{k,\mathcal{T}}^{GR}$ | 6.0000e03 | 6.0674e03 | 6.0000e03 | 6.0674e03 |

TABLE III
SYSTEM-WIDE SETTLEMENT OUTCOMES UNDER DIFFERENT TEMPORAL REPRESENTATIONS

| Quantity | S1 | S2 | S3 | S4 |
|---|---|---|---|---|
| $\Pi_{\mathcal{T}}^{EP}$ ($) | 1.0606e04 | 1.1231e04 | 1.1407e04 | 1.2062e04 |
| $\Pi_{\mathcal{T}}^{OMP}$ ($) | 1.0000e03 | 1.0112e03 | 6.4642e02 | 6.2909e02 |
| $\Pi_{\mathcal{T}}$ ($) | 1.1606e04 | 1.2242e04 | 1.2053e04 | 1.2691e04 |
| Absolute energy imbalance (MWh) | 14.7416 | 0 | 14.7416 | 0 |
| $\sum_{k=1}^{K}\Upsilon_{k,\mathcal{T}}^{ER}-\Pi_{\mathcal{T}}^{EP}$ ($) | 2.5157e02 | 0 | 2.4451e02 | 0 |
| $\Pi_{\mathcal{T},s}^{RMM}$ ($) | 3.6854e02 | 0 | 3.8018e02 | 0 |

Although S4 yields the lowest out-of-market payment among the four scenarios, it does not yield the lowest total customer payment. The S2-S4 comparison holds demand and generation trajectories fixed and isolates the settlement effects of the two assigned prices. The associated price trajectory in S4 raises the in-market energy payment by \$830.76 and reduces out-of-market payment by \$382.14, increasing total customer payment by \$448.62 while leaving modeled generation cost unchanged. This occurs because the price trajectory, over $t$=8.3064-30.6121 min, in S4 is higher than the interval-based price used in S2.

Comparisons across the interval-based and trajectory-based demand representations require an additional qualification. The two representations share the same endpoint power levels but are not energy-equivalent over the hour. In this case, the trajectory-based demand representation contains 25 MWh more cleared energy than the interval-based representation. At the common \$25/MWh interval price used in S1 and S2, this difference alone corresponds to \$625.00 of additional in-market energy payment. Accordingly, raw total-customer-payment differences across the two demand representations should not be interpreted as cost differences for serving the same quantity of energy. Separately, S1 and S3 share the same power-imbalance trajectory but have RMM values of \$368.54 and \$380.18, respectively, because the same imbalance is evaluated using different settlement prices.

In S3, Gen1's selected self-schedule includes additional output during the positive-margin intervals under the trajectory-based price in Fig. 5(f). Relative to the generation reconstruction in Fig. 5(a), this yields a PMLOC of $18.19, as reported in Table II. In S4, PMLOC is zero under the self-scheduling benchmark, so the assumed out-of-market payment consists entirely of EMC, totaling $629.09.

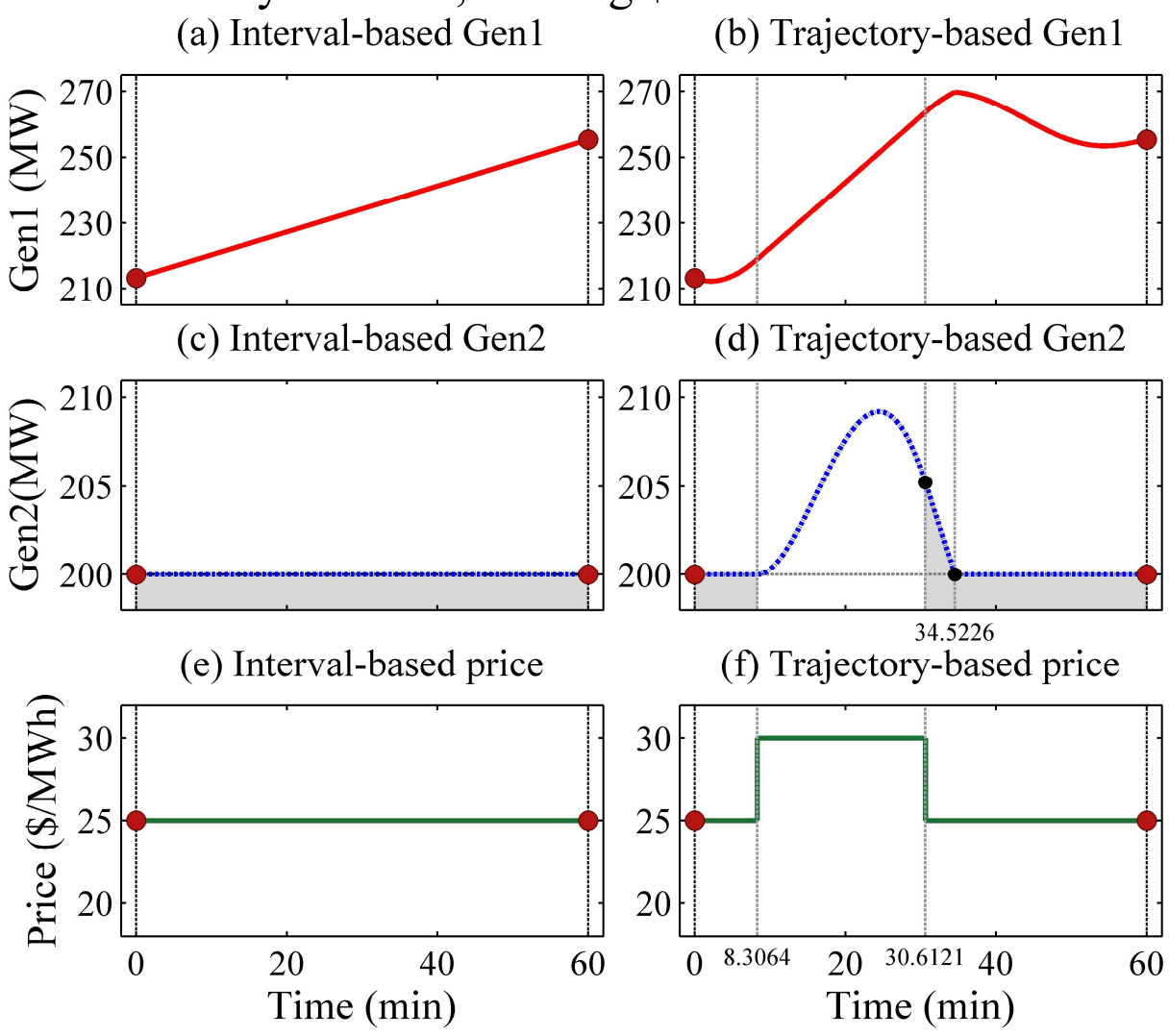

Fig. 5. Generation and price trajectories used for settlement analysis: (a), (c), (e) interval-based representations; (b), (d), (f) trajectory-based representations. The shading in (c) and (d) identifies below-cost production contributing to Gen2's EMC in S1 and S4, respectively.

To identify its timing and physical sources, Gen2's EMC is separated into a minimum-output component and a contribution associated with above-minimum output during below-cost production. In this case, the latter occurs during the ramp-constrained reduction following the price decrease and is identified as the ramp-related component. The price decreases from $30/MWh to $25/MWh at approximately $t$=30.6121 min, while Gen2 remains above its 200-MW minimum output. Its output subsequently decreases under the ramping constraint and reaches the minimum at approximately $t$=34.5226 min, as shown in Fig. 5(d). The above-minimum output during this transition contributes approximately $0.8495 to EMC. The minimum-output component, accumulated over the below-cost periods before $t$=8.3064 min and after $t$=30.6121 min, is approximately $628.24, giving a total EMC of $629.09.

Relative to S1, Gen2's minimum-output EMC in S4 is approximately $371.76 lower because the associated price equals its $30/MWh marginal cost over t=8.3064-30.6121 min. Thus, this component depends on the minimum output level and the magnitude and duration of the price shortfall. By contrast, S2 applies a constant price of $25/MWh to the same generation trajectory as S4. This price remains below Gen2's marginal cost throughout the hour, which makes Gen2's dispatch appear economically unsupported by energy-market revenues alone.

The illustrative 5-bus case shows that balancing the declared demand trajectory does not, by itself, eliminate the supply-side compensation defined in Section IV. In S4, the remaining out-of-market payment consists entirely of EMC, which is separated into minimum-output and ramp-related components under the associated price trajectory. S4 preserves the adopted dispatch-price correspondence and has the lowest out-of-market payment among S1-S4. This result does not imply a lower total customer payment or total generation cost.

## B. *Results in the Modified IEEE RTS-GMLC*

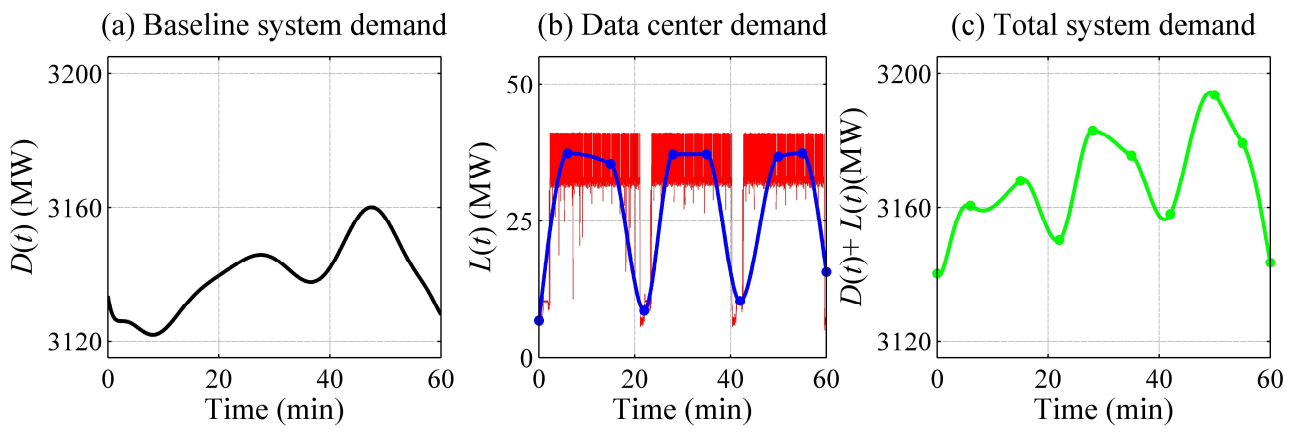

Fig. 6. Continuous-time demand profiles: (a) baseline system demand; (b) scaled workload trace (red) and the fitted declared trajectory used in clearing (blue); and (c) total system demand.

The generating-unit parameters are adapted from the RTS-GMLC system, following the parameterization in [14]. The dispatch and settlement analysis includes 18 generating units. $D(t)$ is constructed by fitting a 12th-degree polynomial to the RTS-GMLC 5-min real-time load data for February 1, 2020, over the one-hour market-clearing horizon. The fitted trajectory is treated as a prescribed input to the deterministic market-clearing problem. An illustrative data center load profile is constructed by scaling a workload-driven computing-power trace from [16] by a factor of 1700. $L(t)$ is then obtained as a continuous piecewise-cubic fit to the scaled profile.

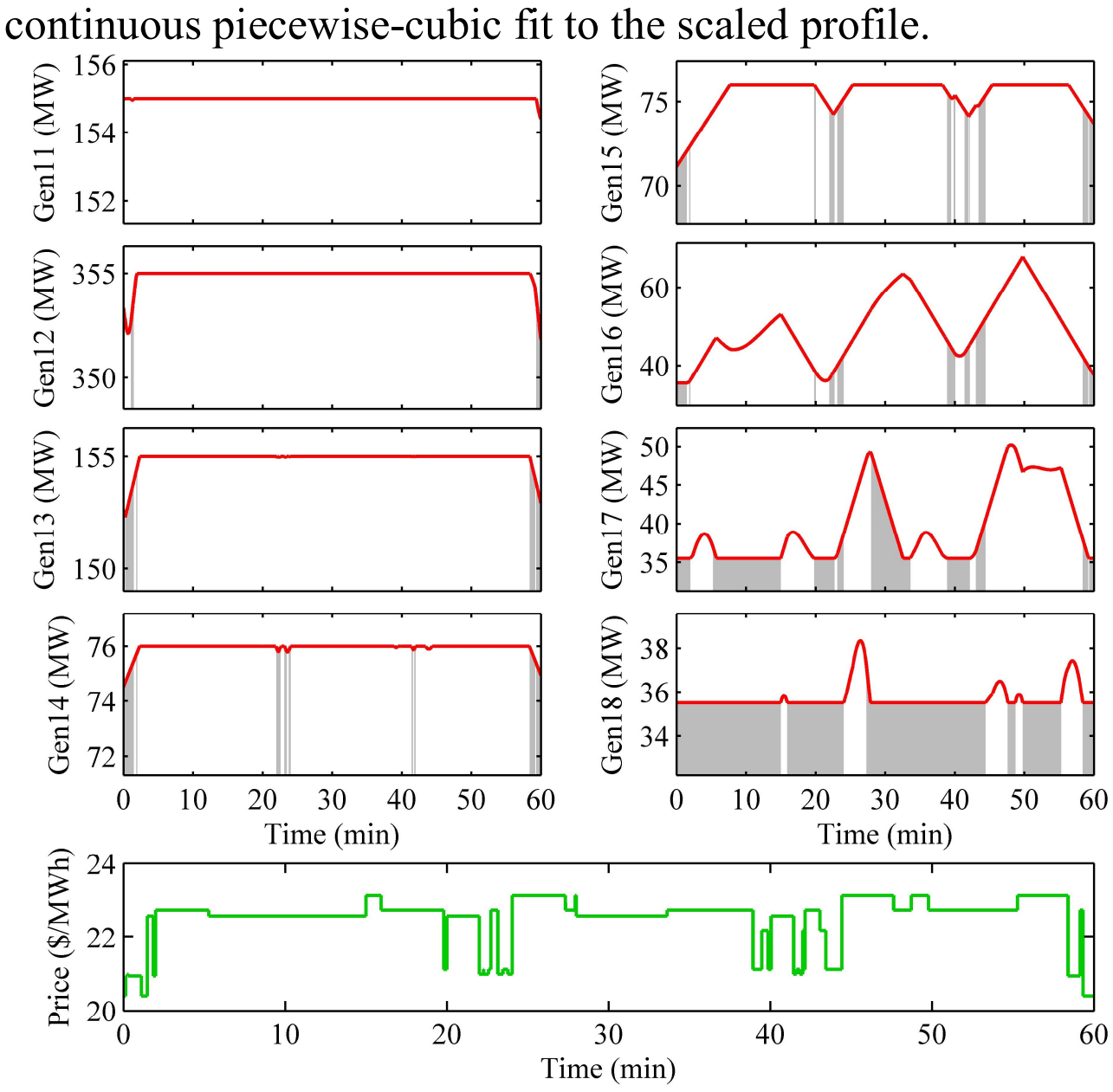

Fig. 7. Generation trajectories of Gen11-Gen18 and the associated price trajectory. Shaded regions identify time intervals during which the market-clearing price is below the unit's marginal cost.

Fig. 6 shows how the constructed workload-driven demand trajectory introduces short-term variations into the total system demand. Fig. 7 presents the generation trajectories of Gen11-Gen18 and the associated price trajectory. The shaded intervals satisfy $\lambda^C(t)<f_k$ and contribute to EMC whenever the cleared output is positive, as defined in (15).

Under the modeled generation operating constraints, a unit may continue producing at or above its minimum output during these below-cost intervals. Gen18 operates at its minimum output for much of the hour, with brief increases in output. Below-cost prices during many of these minimum-output durations contribute to its EMC, which is the largest among the reported units at approximately $21.11. Following the price decrease at approximately $t$=28 min, Gen17 remains above its

35.5-MW minimum output. Its output decreases from approximately 49.0 MW toward the 35.5-MW minimum output following a ramp-limited transition and reaches the minimum at approximately $t$=32.56 min. The above-minimum output maintained during this below-cost transition contributes to the ramp-related component of Gen17's EMC.

Table IV reports ER, PMLOC, and EMC individually for Gen11-Gen18 and in aggregate for Gen1-Gen10. Total EMC is approximately $58.23 and is distributed among Gen12-Gen18, with the largest contribution from Gen18 at approximately $21.11. Total PMLOC is zero.

TABLE IV
GENERATOR-SIDE SETTLEMENT COMPONENTS IN THE MODIFIED IEEE RTS-GMLC SYSTEM.

| Unit | $\Upsilon_{k\mathcal{T}}^{\mathrm{ER}}$ ($) | $\Upsilon_{k\mathcal{T}}^{\mathrm{PMLOC}}$ ($) | $\Upsilon_{k\mathcal{T}}^{\mathrm{EMC}}$ ($) |
|---|---|---|---|
| Gen11 | 3.4927e03 | 0 | 0 |
| Gen12 | 7.9978e03 | 0 | 3.8039e00 |
| Gen13 | 3.4910e03 | 0 | 1.9788e00 |
| Gen14 | 1.7116e03 | 0 | 1.7185e00 |
| Gen15 | 1.6999e03 | 0 | 1.1218e01 |
| Gen16 | 1.1194e03 | 0 | 8.3777e00 |
| Gen17 | 8.9352e02 | 0 | 1.0017e01 |
| Gen18 | 8.0479e02 | 0 | 2.1111e01 |
| Gen1-Gen10 | 5.0184e04 | 0 | 0 |
| Total | 7.1395e04 | 0 | 5.8225e01 |

Overall, the modified IEEE RTS-GMLC case extends the trajectory-based settlement analysis to a larger set of generating units under a constructed workload-driven demand trajectory. The joint representation of generation and prices in Fig. 7 makes the timing of below-cost production explicit and supports interpretation of EMC contributions associated with minimum-output operation and transitions shaped by the ramping constraints. These results illustrate how the temporal distribution and unit-level allocation of the supply-side compensation components can be examined within the adopted clearing and pricing formulation.

## VI. CONCLUSION AND FUTURE WORK

This paper develops a power-trajectory framework for data center participation in day-ahead markets. The declared demand-side power trajectory retains predictable intra-interval evolution and enters clearing as an exogenous input, placing demand representation, dispatch, pricing, and settlement on a common temporal basis. For feasible market-clearing outcomes, continuous-time power balance and satisfaction of the modeled generation operating constraints establish physical temporal consistency. For these outcomes, time-matched settlement using the corresponding price trajectory preserves dispatch-price correspondence and establishes economic temporal consistency within the adopted formulation.

In the 5-bus case, trajectory-based dispatch eliminates representation-induced power imbalance, while supply-side compensation remains. Under the adopted compensation convention, the matched trajectory-based pairing yields the lowest out-of-market payment among the four tested pairings, but not the lowest total customer payment. The modified IEEE RTS-GMLC case illustrates the timing and unit-level distribution of compensation under workload-driven demand variations. Together, these results show that eliminating representation-induced power imbalance does not necessarily eliminate the defined supply-side compensation. The settlement analysis reports RMM separately as a monetary diagnostic and makes explicit how the remaining compensation relates to minimum generation limits, ramping constraints, and associated price transitions.

Future work will incorporate network constraints, unit commitment, reserve procurement, and balancing actions into coordinated day-ahead and real-time clearing and settlement.